\documentclass[prl,reprint,twocolumn,superscriptaddress,amsmath,amssymb,floatfix]{revtex4-2}
\usepackage{soul}
\usepackage{orcidlink}
\usepackage{graphicx}
\usepackage{epstopdf}
\usepackage[utf8]{inputenc}
\usepackage{amsmath,amsthm,amssymb}
\usepackage{latexsym}
\usepackage{bm}
\usepackage{dcolumn}
\usepackage{braket}
\usepackage[normalem]{ulem}
\usepackage{times}
\usepackage{hyperref}
\usepackage[section]{placeins}

\begin{document}
\bibliographystyle{apsrev4-2}

\title{Bogolyubov-Feynman-Bohm quasiparticles in superfluid helium}
\title{End of sound in quantum liquid}
\title{Breakdown of sound quasiparticle in superfluid helium}
\title{Breakdown of sound in superfluid helium}

\author{Marc D. Nichitiu \orcidlink{0000-0001-5330-2676} }
\affiliation{Condensed Matter Physics and Materials Science Division, Brookhaven National Laboratory, Upton, NY 11973, USA}

\author{Craig Brown \orcidlink{0000-0002-9637-9355} }

\affiliation{NIST Center for Neutron Research, National Institute of Standards and Technology, Gaithersburg, MD 20899, USA}


\author{Igor A. Zaliznyak \orcidlink{0000-0002-8548-7924} }
\email{zaliznyak@bnl.gov}
\affiliation{Condensed Matter Physics and Materials Science Division, Brookhaven National Laboratory, Upton, NY 11973, USA}

\begin{abstract}
{
Like elementary particles carry energy and momentum in the Universe, quasiparticles are the elementary carriers of energy and momentum quanta in condensed matter. And, like elementary particles, under certain conditions quasiparticles can be unstable and decay, emitting pairs of less energetic ones. Pitaevskii proposed that such processes exist in superfluid helium, a quantum fluid where the very concept of quasiparticles was borne, and which provided the first spectacular triumph of that concept. Pitaevskii's decays have important consequences, including possible breakdown of a quasiparticle. Here, we present neutron scattering experiments, which provide evidence that such decays explain the collapsing lifetime (strong damping) of higher-energy phonon-roton sound-wave quasiparticles in superfluid helium. This damping develops when helium is pressurized towards crystallization or warmed towards approaching the superfluid transition. Our results resolve a number of puzzles posed by previous experiments and reveal the ubiquity of quasiparticle decays and their importance for understanding quantum matter.
}


\end{abstract}

\date{\today}

\maketitle
\newpage

The quasiparticle concept is a cornerstone of our understanding of many-body atomic systems that make up materials around us. Heat, sound, electric current, and their inter-conversion in materials which underpin a broad range of technologies, can all be understood as being carried by elementary excitations, the quasiparticles. These elementary excitations were devised by Landau to describe the properties of superfluid helium isotope $^{4}$He, a quantum liquid with zero viscosity which can flow without any friction \cite{Kapitza_1938}, and led to a triumph in our understanding of quantum condensed matter \cite{Landau_JPhysUSSR1941,Landau_PhysRev1941,Landau_JPhysUSSR1947}. By postulating the energy-momentum relation, $\epsilon(Q)$, of phonon-roton quasiparticles -- sound waves carrying energy and momentum in superfluid $^{4}$He, Landau very accurately explained essentially all of its experimentally observed properties  ($Q = p/\hbar$ is the wave vector of the corresponding wave with wavelength $\lambda = 2 \pi/Q$, $p$ is momentum, $\hbar$ is Plank's constant). For quasiparticles, this $\epsilon(Q)$ dispersion replaces the famous Einstein's relation, $\epsilon(p) = \sqrt{(pc)^2 + (mc^2)^2}$ ($c$ is velocity of light and $m$ is the particle mass at rest), or its non-relativistic Newtonian version, $\epsilon(p) = p^2/2m$, which describe elementary particles.

\begin{figure}[b!h!]
\includegraphics[width=1.\linewidth]{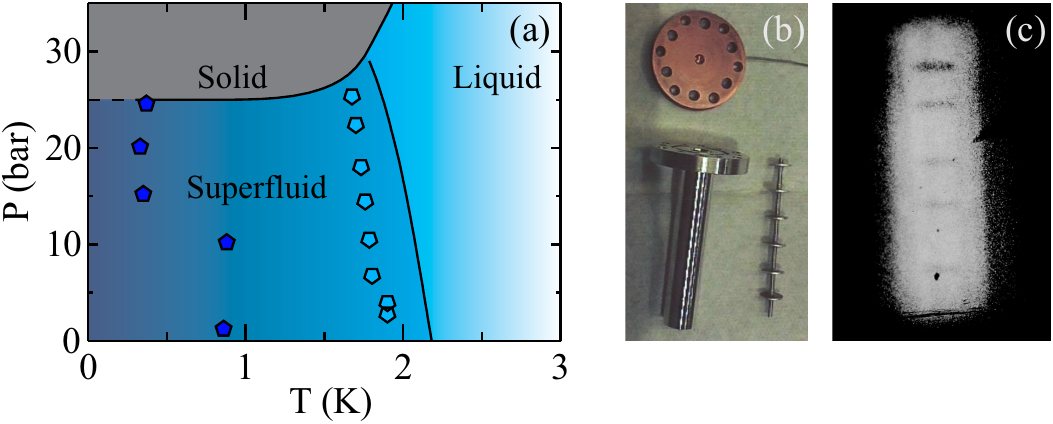}
\caption{{\bf The (P, T) phase diagram of superfluid $^4$He with locations of our INS measurements. } (a) lines indicate superfluid, normal liquid, and solid phase boundaries; dark-filled upward-pointing pentagons represent the low-temperature measurements with incident neutron energy $E_i = 3.55$~meV shown in Fig.~\ref{Fig1:neutron_maps_lowT}, and light-filled downward pentagons represent measurements closer to the superfluid-to-normal-liquid transition with $E_i = 3.27$~meV shown in Fig.~\ref{Fig2:neutron_maps_highT}. (b) the Al sample cell ($0.95$~cm inner, $1.25$~cm outer diameter) used in our measurements, with Al rod with $\approx 1$~cm spaced Cd dividers and copper top flange with the $\approx 0.15$~cm diameter input capillary. (c) transmission neutron radiograph of the (empty) assembled sample cell, with shades from neutron-absorbing Cd dividers visible. The resulting small, $\approx 0.7$~cm$^3$ scattering volume of each sub-cell markedly reduces parasitic double scattering effects compared to previously reported measurements \cite{Fak_JLTP1992,Fak_JLTP1998,Gibbs_JPCM1999,Beauvois_PRB2016,Godfrin_PRB2021}. }
\label{Fig0:PhaseDiagram}
\end{figure}

A detailed theory of viscosity behavior in superfluid helium was subsequently developed by Landau and Khalatnikov (LK) based upon the idea that transport phenomena can be described in terms of collisions between the quasiparticles, which form a nearly ideal gas \cite{Haar_Landau_collected_papers_book1965,Khalatnikov_book}. The resulting phonon-roton transport theory provided very good agreement with the experimental values of the viscosity coefficient at low temperatures. A notable deviation observed in a temperature region near the superfluid transition [lambda point, $T_{\lambda} \approx 2.19$~K at atmospheric pressure $\approx 1$~bar, Fig.~\ref{Fig0:PhaseDiagram}(a)] was ascribed to the failure of the approximation where phonon and roton quasiparticles are treated as nearly ideal gases. Here, we show that it is the quasiparticle breakdown processes that are at the origin of the observed deviation.

That Landau’s guess for $\epsilon(Q)$ turned out to be remarkably accurate was confirmed by inelastic neutron scattering (INS), a technique
that allows to directly detect quasiparticles \cite{HenshawWoods_PhysRev1961,WoodsCowley_RepProgPhys1973,Dietrich_etal_Passell_PRA1972,Graf_etal_Passell_PRA1974,Woods_JPhysC1977,Stirling_PRB1990,Andersen_JPCM1994,Fak_JLTP1992,Fak_JLTP1998,Montfrooij_JLTP1997,Gibbs_JPCM1999,Montfrooij_arXiv2006,Beauvois_PRB2018,Godfrin_PRB2021}. 
By measuring the probability for a neutron passing through a superfluid helium (or another material) to scatter losing some of its energy and momentum, one can experimentally determine the energy-momentum relationship of the quasiparticles that are created as a result. Recent progress in neutron scattering technology allows to conduct such measurements with exceptional precision \cite{Beauvois_PRB2018,Godfrin_PRB2021}. It was shown \cite{Godfrin_PRB2021} that using the precise INS measurements of quasiparticle dispersion in superfluid $^4$He as an input to Landau theory provides exceptionally accurate description of specific heat and other thermodynamic properties at low temperatures, while notable discrepancies are present within about $0.5$~K below the superfluid transition. By exploring in detail the behavior of phonon-roton quasiparticle in this region in our INS experiments (Fig.~\ref{Fig0:PhaseDiagram}), we show that the observed failure of theoretical description is rooted not in the non-ideal nature of phonon and roton gases implied by the most simple version of the Landau theory used in \cite{Godfrin_PRB2021}, but, in fact, in the failure of the quasiparticle description at its core borne by the quasiparticle decay processes.

\begin{figure}[t!]
\includegraphics[width=1.\linewidth]{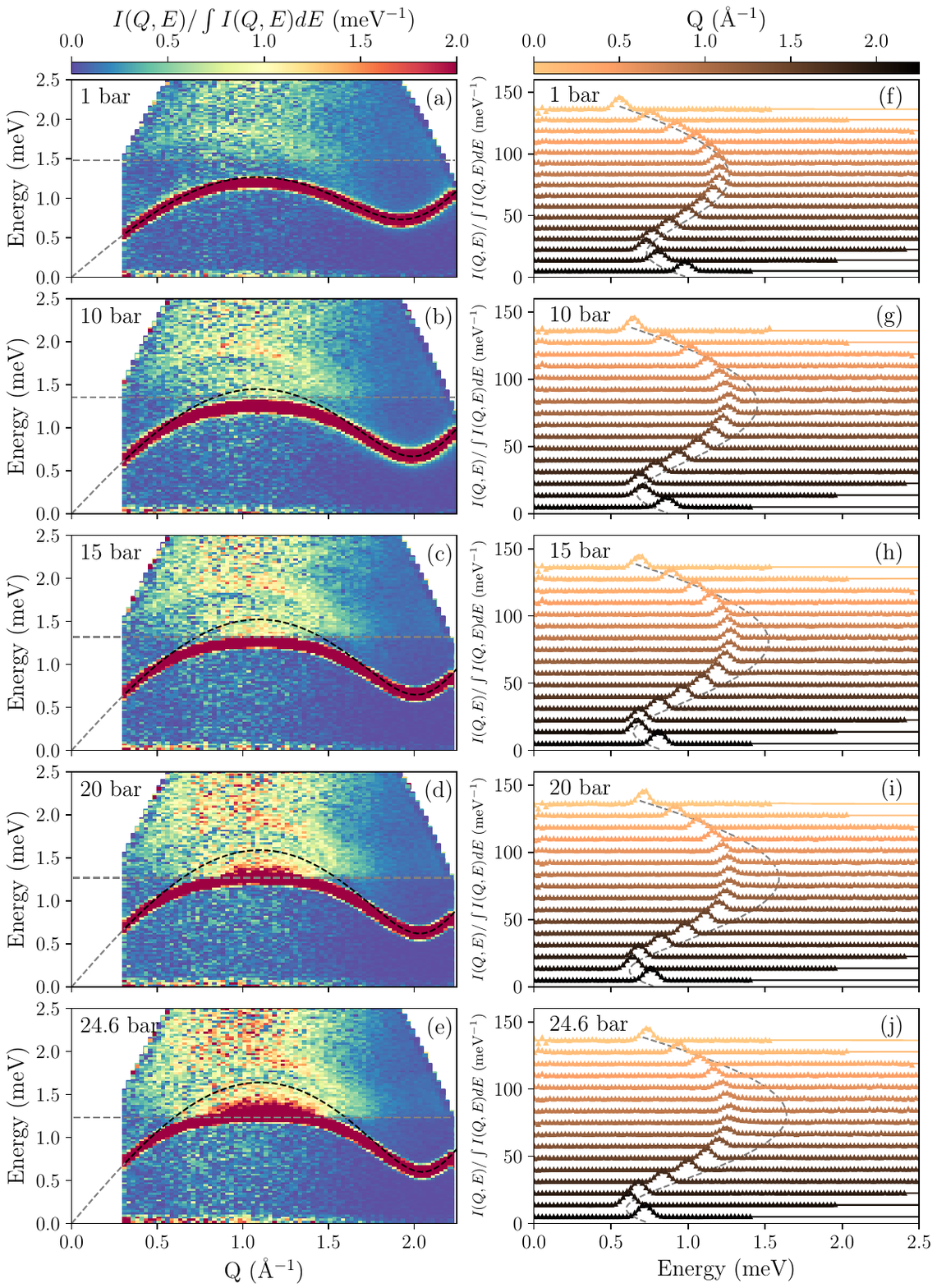}
\caption{{\bf Phonon-roton quasiparticle in superfluid helium-4.} (a)--(e) color contour maps of the spectral density of the measured neutron scattering intensity at different pressures, $P = $~1.22(3), 10.19(3), 15.20(3), 20.11(2), and 24.57(4) bar (top to bottom), tracking the quasiparticle dispersion, $\epsilon(Q)$. 
The dashed curve is the fitted Bogolyubov dispersion \cite{Feynman_PhysRev1954,BogolyubovZubarev_JETP1955,BohmSalt_RMP1967,Sunakawa_ProgTheorPhys1969,IsiharaSamulski_PRB1977,ZaliznyakTranquada_2014} without accounting for multi-particle interactions, the horizontal dashed line marks the decay threshold energy, $2\Delta$. (f)--(j) selected constant-$Q$ cuts of the corresponding {spectral density of the} measured neutron intensity with fits to resolution-corrected DHO lineshape. The width of the peak is consistent with the instrument resolution ($\Delta E_{res} \approx 0.1$~meV), with a visible deviation at the highest pressures near the top of dispersion, where pair decays become active. }
\label{Fig1:neutron_maps_lowT}
\vspace{-0.25in}
\end{figure}

Color contour plots in Figure~\ref{Fig1:neutron_maps_lowT}~(a)-(e) show spectral density of INS intensity obtained in our measurements at different pressures and at low temperatures within the superfluid phase (Fig.~\ref{Fig0:PhaseDiagram}, see Figure~S1 for the measured neutron intensity). The intense curvy line traces the phonon-roton quasiparticle dispersion (linear rise at small $Q$ is the phonon part, while the minimum near $Q \approx 1.9$~\AA$^{-1}$, following Landau, is called roton). The narrow peak of the measured scattering intensity in Fig.~\ref{Fig1:neutron_maps_lowT}(f)-(l), which reveals the quasiparticle, has width which is approximately consistent with the experimental energy resolution of our measurement (see also \cite{Supplementary} and discussion below). Zero, or small intrinsic energy width of the INS peak indicates infinite, or very long quasiparticle lifetime. Ideally, quasiparticles with infinite lifetime, $\tau$, describe stationary excited states of the system that are the eigenstates of energy and momentum with the eigenvalues in one-to-one correspondence forming unique pairs, $(\epsilon, Q)$, which determine the quasiparticle dispersion, $\epsilon(Q)$. INS measures the probability of different energy-momentum excited states of the system and in such an ideal case, $\tau = \infty$, the probability distribution is Dirac's delta function, $\delta (\epsilon - \epsilon(Q))$, of zero width, $\Gamma = \hbar/\tau = 0$ (in practice, in an INS experiment the measured distribution is broadened by a finite instrument resolution resulting from less than perfect discrimination between different energies and wave vectors). In such case, the knowledge of the quasiparticle dispersion, $\epsilon(Q)$, is sufficient to construct the systems's partition function and accurately describe all of its thermal properties, which explains the power of the quasiparticle concept.

While considering quasiparticle collisions and applying Boltzmann transport theory allows {us} to obtain a quantitative description of transport phenomena in superfluid helium, such as convection and viscosity \cite{Haar_Landau_collected_papers_book1965,Khalatnikov_book}, collisions also shorten quasiparticle lifetime. Like for elementary particles, collisions can change quasiparticles' $(\epsilon, Q)$ identities, which means that in the presence of other excited states{,} each elementary excitation acquires a finite lifetime. 
The more quasiparticles are excited with the increasing temperature, storing the system's thermal energy, the shorter their lifetime {becomes} due to collisions, which {increase in frequency}. {Landau and Khalatnikov (LK)} obtained an accurate description of these effects, where the roton peak width is proportional to the number of thermally excited rotons, $\Gamma \sim \sqrt{T} e^{-\frac{\Delta(T)}{k_B T}}$ ($\Delta$ is the minimum energy of the roton, Fig.~\ref{Fig1:neutron_maps_lowT}, which depends on $P$ and $T$). This was checked in a number of INS experiments and was shown to work well up to $\approx 1.5$~K \cite{Andersen_PRL1996,Fak_PRL2012}. Apparent deviations from {LK} behaviour of $\Gamma$ at higher temperatures were considered to reflect the inadequacy of the experimental model in extracting quasiparticle parameters rather than any inherent inaccuracy of the LK model \cite{Gibbs_JPCM1999}.

At zero temperature, there are no collisions that would limit the quasiparticle lifetime because there are no thermally excited quasiparticles. Hence, according to LK, $\Gamma (T) = 0$ at $T = 0$, which means that the measured width of the INS spectrum should be resolution limited. However, quasiparticles, like some elementary particles, can be unstable with respect to decays if these are allowed by quantum-mechanical conservation laws. Such spontaneous decays can lead not only to a finite lifetime, but {also} to a complete disappearance of the quasiparticle states. Landau conjectured that the quasiparticle spectrum in superfluid $^4$He could terminate at large $Q$ where its energy increases such that decays into roton pairs become kinematically allowed, {i.e.} the quasiparticle dispersion enters the energy range of the continuum of two-roton states,  $\epsilon (Q) \geq 2\Delta$. An elegant theory of this phenomenon was developed by Pitaevskii to whom the problem was posed \cite{Pitaevskii_JETP1959,LandauLifshitz_Vol9_1980}. Not only did {this theory} predict the spectrum end point, $Q_c$, but {it} also explained the downward bending of the dispersion on approaching the $Q_c$, a puzzling behavior observed in experiment \cite{WoodsCowley_RepProgPhys1973,Dietrich_etal_Passell_PRA1972,Graf_etal_Passell_PRA1974,Woods_JPhysC1977,Stirling_PRB1990,Andersen_JPCM1994,Fak_JLTP1992,Fak_JLTP1998,Montfrooij_JLTP1997,Gibbs_JPCM1999,Montfrooij_arXiv2006,Beauvois_PRB2018,Godfrin_PRB2021,Glyde_EPL1998}. While similar effects of spontaneous quasiparticle decays have been also observed by INS in quantum magnets \cite{StoneZaliznyak_etal_Nature2006,ZhitomirskyChernyshev_RMP2013}, revealing these to be a ubiquitous property of quantum matter, the unambiguous experimental identification of $Q_c$ in superfluid helium still remains a challenge \cite{Beauvois_PRB2018,Godfrin_PRB2021}.

While the spectrum termination point at high $Q$ is outside the range of our present measurements, the effects of decay interactions transpire in our data at higher pressures (Fig.~\ref{Fig1:neutron_maps_lowT}). As the roton gap, $\Delta(P)$, decreases with the pressure increasing towards crystallization (which for $T \approx 0$ occurs at $P \approx 25$~bar, Fig.~\ref{Fig0:PhaseDiagram}) \cite{Dietrich_etal_Passell_PRA1972,Graf_etal_Passell_PRA1974}, the threshold for the onset of two-roton states also decreases. When this threshold, $2 \Delta(P)$, approaches the local dispersion maximum near $Q_m \approx 1.2$~\AA$^{-1}$ (quasiparticles in this range are conventionally called ``maxon''), the interaction effects first lead to ``squaring'' of the dispersion, which becomes increasingly clear for $P \geq 15$~bar. This energy-level-repulsion effect between the quasiparticle and the two-roton continuum, similar to that predicted by Pitaevskii, is prominently displayed by the increasing (with pressure) discrepancy between the measured $\epsilon(Q)$ and the fitted Bogolyubov dispersion for non-interacting quasiparticles \cite{Feynman_PhysRev1954,BogolyubovZubarev_JETP1955,BohmSalt_RMP1967,Sunakawa_ProgTheorPhys1969,IsiharaSamulski_PRB1977,ZaliznyakTranquada_2014}\textcolor{blue}{,} which accurately describes the lower-energy region of the phonon-roton dispersion where the effects of interaction are small (dashed curve in Fig~\ref{Fig1:neutron_maps_lowT}, see also Figs.~S1, S5).

\begin{figure}[t!h!]
\includegraphics[width=1.\linewidth]{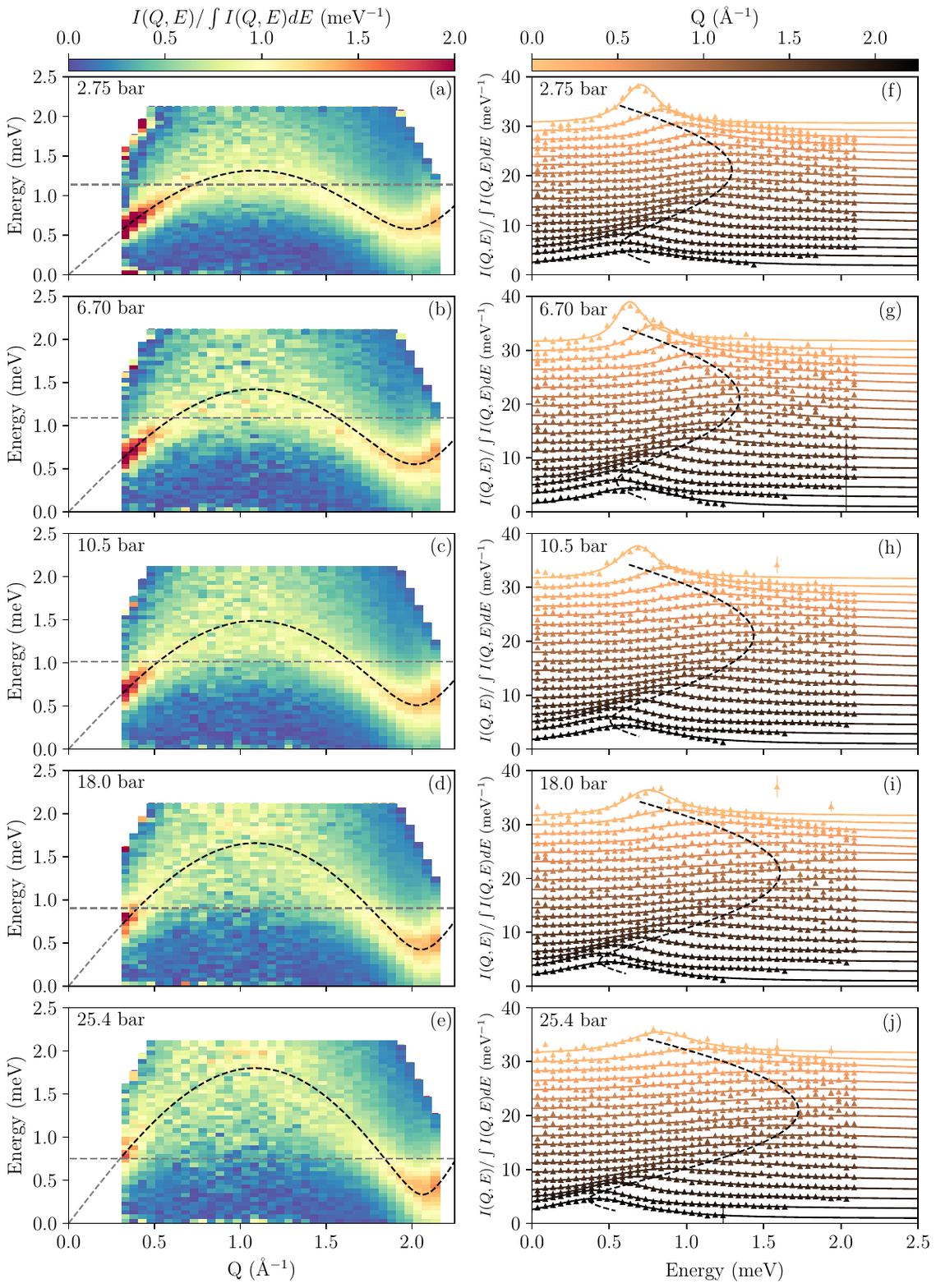}
\caption{{\bf Breakdown of phonon-roton sound-wave quasiparticle near the crystallization and superfluid transition.} (a)--(e) color contour maps of the spectral density of the measured neutron scattering intensity at pressures $P = $~2.75(3), 6.76(8), 10.50(3), 18.03(3), and 25.36(3) bar (top to bottom), at temperatures in $1.7$~K to $1.9$~K range, as shown in Fig.~\ref{Fig0:PhaseDiagram}. As in Fig.~\ref{Fig1:neutron_maps_lowT}, the dashed curve is the fitted Bogolyubov dispersion without accounting for multi-particle interactions, the horizontal dashed line marks the decay threshold energy, $2\Delta$. (f)--(j) selected constant-$Q$ cuts of the corresponding {spectral density of the} measured neutron intensity with fits to resolution-corrected DHO lineshape. The substantial DHO width reflects short quasiparticle lifetime, which decreases markedly above the decay threshold where the spectral density gets extremely blurred. }
\label{Fig2:neutron_maps_highT}
\vspace{-0.25in}
\end{figure}

At the highest pressure of our measurements, $P = 24.6$~bar, the maxon intensity notably decreases and an indication of finite lifetime (finite peak width) appear near $Q_m$, extending similar observation of Ref.~\cite{Godfrin_PRB2021}. With the further decrease of the two-roton threshold energy at higher pressure, the quasiparticle spectrum can be expected to terminate at two wave vectors, $Q_{c 1}$ and $Q_{c 2}$ around $Q_m$, opening an entire region of excited states in the $[Q_{c 1}, Q_{c 2}]$ range to higher-energy excitations. Strikingly, this does not happen. Instead, with a tiny increase in pressure to $\approx 25$~bar superfluid $^4$He solidifies in a first-order phase transition. This occurs well before the roton energy softens to zero, as could be expected for a soft-mode second-order transition. Hence, it appears as if the (avoided) quasiparticle breakdown is actually the cause of the observed ``premature'' crystallization, a well-known puzzling behavior of the superfluid $^4$He. Interestingly, this observation can be understood from a simple quantum-mechanical argument. Superfluid $^4$He at ambient pressure remaining liquid down to an absolute zero temperature hinges on a fine balance between the energy of zero-point quantum motion of the liquid and the solid phases. In a superfluid, zero-point energy is determined by the quasiparticle dispersion, $E_{0} = \sum_{Q} {\frac{1}{2} \epsilon(Q)}$. When the quasiparticles break down between $Q_{c 1}$ and $Q_{c 2}$, higher-energy excitations contribute to zero-point motion and its energy increases to become larger than that of a solid, causing crystallization.

\begin{figure}[t!]
\includegraphics[width=1.\linewidth]{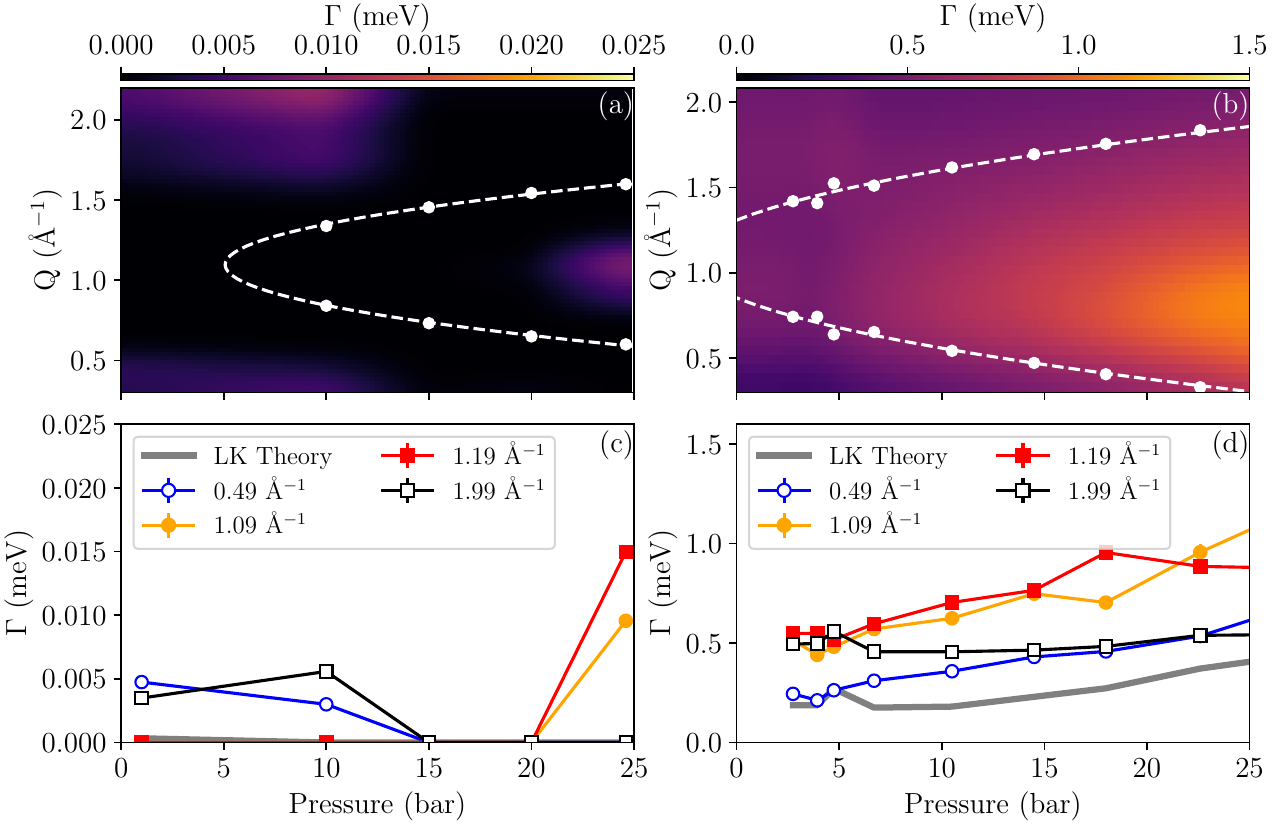}
\caption{{\bf  The quasiparticle width and the breakdown region.} Color contour map of the DHO half width at half maximum (HWHM), $\Gamma$, which parameterizes the quasiparticle lifetime, $\tau \sim h/\Gamma$, obtained by interpolation of the fit results as a function of pressure, (a) for the low-temperature data of Fig.~\ref{Fig1:neutron_maps_lowT} and (c) for the data of Fig.~\ref{Fig2:neutron_maps_highT} with pronounced decays. The solid symbols with the parabolic fit (dashed line) show the boundary of the pressure-dependent quasiparticle breakdown region of momenta, $[Q_{c 1}, Q_{c 2}]$, where decays are allowed for non-interacting quasiparticles with the fitted Bogolyubov dispersion \cite{Feynman_PhysRev1954,BogolyubovZubarev_JETP1955,BohmSalt_RMP1967,Sunakawa_ProgTheorPhys1969,IsiharaSamulski_PRB1977,ZaliznyakTranquada_2014} of Figs.~\ref{Fig1:neutron_maps_lowT},~\ref{Fig2:neutron_maps_highT}(a)--(e) (see also Figs.~S1--S7). The pressure dependence of $\Gamma$ for typical wave vectors in the phonon (open circles), maxon (filled circles and squares) and roton (open squares) regions (b) for the data in (a) and (d) for the data in (c). The grey line shows $\Gamma$ obtained from LK theory \cite{Andersen_PRL1996,Fak_PRL2012}. The experimental $\Gamma$ obtained using a single-component DHO fit is somewhat over-estimated by the inclusion of multi-particle states in the fitted intensity, however, its variation with pressure and $Q$ adequately exposes the physics of quasiaprticle decays (see also \cite{Supplementary}). Error bars in all figures show one standard deviation and where not visible are smaller than the symbol size.
}
\label{Fig3:Gamma}
\vspace{-0.25in}
\end{figure}

With temperature increasing towards the $\lambda$ point, the roton energy and the two-roton threshold further decrease \cite{Dietrich_etal_Passell_PRA1972,Montfrooij_JLTP1997,Gibbs_JPCM1999}. For temperatures $\sim 0.8 T_\lambda$ of our INS measurements presented in Fig.~\ref{Fig2:neutron_maps_highT} ($1.7$~K to $1.9$~K, Fig.~\ref{Fig0:PhaseDiagram}), even at low pressure of $2.75$~bar the two-roton threshold intercepts the phonon-roton dispersion, allowing decay processes within a finite $Q$-range around $Q_m$ [Fig.~\ref{Fig2:neutron_maps_highT}(a)]. While the entire quasiparticle spectrum already has substantial thermal width, $\lesssim 0.5$~meV, because of the finite LK lifetime due to collisions at these temperatures \cite{Dietrich_etal_Passell_PRA1972,Gibbs_JPCM1999} (see also Fig.~\ref{Fig3:Gamma}), there is a marked additional blurring near the top of dispersion due to decays. This effect is most clearly seen at low $Q$, where for energies below the two-roton threshold [horizontal dashed line in Fig.~\ref{Fig2:neutron_maps_highT}(a)-(e)], the phonon quasiparticle in the linear part of the dispersion presents a well-defined peak in the measured INS intensity at each $Q < Q_{c1}$, only broadened by a finite lifetime [bright streak at low $Q$ in Fig.~\ref{Fig2:neutron_maps_highT}(a)-(e)]. The peak broadens dramatically for energies above the threshold, $\epsilon > 2 \Delta$, revealing the effect of decays. The same is the situation in the roton region, where an LK collision-lifetime-broadened quasiparticle exists below the $2 \Delta$ threshold, for $Q > Q_{c2}$. We note that at finite temperature quasiparticle breakdown for $Q \in [Q_{c1},Q_{c2}]$ does not lead to crystallization because the liquid state is entropically stabilized. Albeit access to higher-energy excited states does increase the system's internal energy, $E$, it also adds to the entropy, $S$, whose contribution at finite $T$ lowers the free energy of the system, $E - k_B T \ln S$.

Comparing panels (a) through (e) of Fig.~\ref{Fig2:neutron_maps_highT}, we observe that the decay region expands as pressure increases towards crystallization, blurring an increasingly wider part of the quasiparticle dispersion around $Q_m$. In this ever-increasing $Q$-range, the quasiparticle instability to decays invalidates the LK type theoretical approach to describing transport and thermal phenomena in superfluid $^4$He in terms of quasiparticles and their collisions. This observation explains previously reported discrepancy, growing at higher temperatures and pressures, between the LK theory and INS experiment \cite{Dietrich_etal_Passell_PRA1972,Graf_etal_Passell_PRA1974,Montfrooij_JLTP1997,Gibbs_JPCM1999}.

In the presence of a finite lifetime, the quasiparticle spectral function measured by INS transforms from a Dirac delta function for $\tau = \infty$ to that of a damped harmonic oscillator (DHO), $I(\epsilon, Q) \sim \frac{2 \Gamma \epsilon}{\left(\epsilon^2 - (\epsilon^2(Q)+\Gamma^2) \right)^2 + (2 \Gamma \epsilon)^2}$ (for under-damped case, this expression is equivalent to the difference of two Lorentzian functions centered at $\pm \epsilon(Q)$ and with full width at half maximum (FWHM), $2\Gamma = 2\hbar/\tau$, \cite{Zaliznyak_PRB1994}; at finite $T$, it is also weighted by the detailed balance factor, \cite{Gibbs_JPCM1999,ZaliznyakTranquada_2014}). We therefore quantify the effects of quasiparticle spectrum broadening by fitting the measured INS intensity at each $Q$ to the DHO response with the quasiparticle energy, $\epsilon(Q)$, intensity, $I(Q)$, and $Q$-dependent width, $\Gamma(Q)$, as parameters. The corresponding fits are shown by solid lines in panels (f)-(j) of Figures \ref{Fig1:neutron_maps_lowT} and \ref{Fig2:neutron_maps_highT} and the obtained width, $\Gamma(Q)$, is presented in Figure \ref{Fig3:Gamma}.

For the low-temperature data of Fig.~\ref{Fig1:neutron_maps_lowT}, there is a small but discernible width, $\sim 0.01$~meV, for the $1$~bar and $10$~bar data measured at $T \approx 0.9$~K, which is larger than the LK collisional broadening and probably indicates decays into phonon pairs such as discussed in \cite{Godfrin_PRB2021} [Fig.~\ref{Fig3:Gamma}(a),(b)]. For the $15$~bar and $20$~bar, $\approx 0.35$~K data, there is no discernible broadening, consistent with LK and with the pressure-induced stability of the phonon spectrum \cite{Godfrin_PRB2021} (except for a small effect near $Q_m$ at $20$~bar, indicative of the onset of decays into roton pairs). At $24.6$~bar, also measured at $\approx 0.35$~K, fitting reveals noticeable width, $\Gamma > 0.02$~meV, exceeding that seen at $0.9$~K, for wave vectors near $Q_m$, clearly indicating the effect of quasiparticle decays, which can also be identified in Fig.~\ref{Fig1:neutron_maps_lowT}(e) and Fig.~S1(j). The symbols with fitted parabolic dashed line in Fig.~\ref{Fig3:Gamma}(a) show the pressure-dependent momentum region where decays are allowed for non-interacting quasiparticles with the fitted Bogolyubov dispersion \cite{Feynman_PhysRev1954,BogolyubovZubarev_JETP1955,BohmSalt_RMP1967,Sunakawa_ProgTheorPhys1969,IsiharaSamulski_PRB1977,ZaliznyakTranquada_2014} shown in Fig.~\ref{Fig1:neutron_maps_lowT}(a)-(e). In reality, the interaction-induced decrease of dispersion maximum pushes this region to higher pressures, $\gtrsim 20 $~bar.

The decay region explodes at higher temperatures, as the roton energy decreases, Fig.~\ref{Fig3:Gamma}(c). The two-roton decays add substantially, up to $\gtrsim 100\%$ for the temperatures $1.7$~K to $1.9$~K we measured, to the LK collisional thermal damping, which, albeit already large, only dominates at low energies, $E \lesssim 2\Delta$ [Fig.~\ref{Fig3:Gamma}(d)]. It is therefore not surprising that a quasiparticle transport theory\textcolor{blue}{,} which only accounts for collisions and neglects decays could diverge from experiment at temperatures near the superfluid transition and at pressures close to crystallization where the roton gap becomes small and the decay region is large. 

Superfluid helium presents the standard model of quasiparticle physics in quantum matter \cite{Haar_Landau_collected_papers_book1965,LandauLifshitz_Vol9_1980,Feynman_PhysRev1954}. An understanding and accurate description of the quasiparticles in helium has been foundational for the development of theories of many-body quantum states and is fundamental for the progress in our ability to describe and control quantum systems of Bose particles, from trapped atoms to quantum magnets \cite{StoneZaliznyak_etal_Nature2006,ZhitomirskyChernyshev_RMP2013}. Here, we report experimental observation of an instability towards pair decays leading to breakdown of phonon-roton sound wave, an important aspect of quasiparticle behavior in superfluid helium {that} has been predicted long time ago. Our present results provide a much-needed completion for the standard model of quantum condensed matter, uncovering the origin of the remaining discrepancies between theory and experiment and unveiling an unusual route to zero-temperature crystallization, which resolves a long-standing puzzle.

\begin{acknowledgments}
We gratefully acknowledge the invaluable technical assistance from J. Leao and NCNR staff. IZ is indebted to the late Larry Passell for sharing his wisdom and providing critical advice concerning the INS measurements of liquid helium. We are also grateful to W. Montfrooij, A. Shytov, M. Zhitomirsky, A. Tkachenko, and A. Abanov for valuable discussions. This work at Brookhaven National Laboratory was supported by Office of Basic Energy Sciences (BES), Division of Materials Sciences and Engineering,  U.S. Department of Energy (DOE), under contract DE-SC0012704.  \\
\end{acknowledgments}

\pagebreak

\hypersetup{pageanchor=false}
\renewcommand{\thepage}{S\arabic{page}}
\setcounter{page}{1}
\renewcommand{\theequation}{S\arabic{equation}}
\setcounter{equation}{0}
\renewcommand{\thefigure}{S\arabic{figure}}
\setcounter{figure}{0}

\begin{widetext}

\section*{Supplementary Information}

\begin{center}
{\bf Breakdown of sound in superfluid helium } \\
M. D. Nichitiu \orcidlink{0000-0001-5330-2676}, C. Brown \orcidlink{0000-0002-9637-9355}, and I.~A.~Zaliznyak \orcidlink{0000-0002-8548-7924} \\
\bigskip
correspondence to: zaliznyak@bnl.gov
\end{center}

\bigskip
\noindent{\bf This PDF file includes:}\\
Supplementary Text\\
Supplementary Figures S1-S8\\

\subsection{Experimental procedure and details}
\label{Methods}
Helium gas under controlled pressure was condensed into a cylindrical can made of Aluminum 7075-T6 alloy through a $1/16"$ stainless steel capillary\textcolor{blue}{,} soldered into a $0.25"$ thick top copper flange sealing the can and attached to the cold finger of a $^3$He pumped cryostat with the base temperature of $0.35$~K. In order to reduce parasitic double scattering effects where neutron is scattered twice while traversing the sample, we followed the advice of L. Passell based on the experience with previous measurements \cite{Dietrich_etal_Passell_PRA1972,Graf_etal_Passell_PRA1974} and chose sample cell with the small inner diameter, $0.375"$, and with the wall thickness $0.06"$, Fig.~\ref{Fig0:PhaseDiagram}(b). A threaded Al rod with $\approx 1$~cm spaced neutron-absorbing Cd disks matching the inner diameter of the cell was used to reduce sample scattering volume in the vertical direction, resulting in small, $\approx 0.7$~cm$^3$ volume of each sub-cell and markedly reducing parasitic double scattering effects compared to previously reported measurements \cite{Fak_JLTP1992,Fak_JLTP1998,Gibbs_JPCM1999,Beauvois_PRB2016,Godfrin_PRB2021}. The total height of the sample cell illuminated by neutron beam in our experiments was $\approx 6$~cm [Fig.~\ref{Fig0:PhaseDiagram}(b),(c)]. The background (BG) scattering from the empty cell was measured at the end of each experiment and subtracted from all data. The high quality of BG subtraction (Figs.~\ref{Fig1:neutron_maps_lowT}, S2) indicates high transmission through the sample and validates neglecting the double scattering effects.

The time-of-flight neutron scattering measurements were performed at the Disk Chopper Spectrometer (DCS), NIST Center for Neutron Research (NCNR). For low-temperature measurements shown in Fig.~\ref{Fig1:neutron_maps_lowT}, the incident neutron energy was set to $E_i = 3.55$~meV ($\lambda = 4.8$~\textrm{\AA}); for measurements closer to the superfluid-to-normal-liquid transition shown in Fig.~\ref{Fig2:neutron_maps_highT}, $E_i = 3.27$~meV ($\lambda = 5.0$~\textrm{\AA}) was used. In both cases coarse resolution chopper settings were used to gain intensity; this resulted in elastic energy resolution full width at half maximum (FWHM) $\approx 0.1$~meV (as determined by fitting the scattering from a standard Vanadium sample).

\subsection{Data presentation and additional data}
\label{AdditionalData}
The measured neutron intensity, $I(Q, E)$, was re-histogrammed from an instrument-native detector histograms onto a rectangular grid in $(Q, E)$ with the step $\Delta Q = 0.025$\AA$^{-1}$ and $\Delta E = 0.025$~meV in wave vector and energy, respectively (we explored different bin sizes for the grid to optimize the balance between energy and wave vector resolution and the statistical error of the histogrammed intensity). The $(Q, E)$ color maps and the corresponding line cuts in Fig.~\ref{Fig1:neutron_maps_lowT} and \ref{Fig2:neutron_maps_highT} of the main text show the spectral density of the measured intensity, $f(Q, E)$, obtained from energy cuts at each wave vector, $Q_i$, by normalizing the intensity to the integral intensity of the cut,
\begin{equation}
f(Q_i, E) = I(Q_i, E)/\int{I(Q_i, E)dE} =  I(Q_i, E)/\sum_j{I(Q_i, E_j)\Delta E},
\label{SpectralDensity}
\end{equation}
where the integral was evaluated via numerical summation of the measured intensity weighted by the corresponding energy bin size. The normalization of spectral function removes the intensity dependence on the structure factor, $S(Q) = \int S(Q, E) dE$, which weights the measured intensity to be largest near the roton position and vanish for $Q \rightarrow 0$.

Figure \ref{FigS1:additional_maps_lowT} presents the measured INS intensity [panels (a)--(e)] with the corresponding line cuts [panels (f)--(j)] for the low-temperature, $T < 0.9$~K data, from which the spectral function presented in Fig.~\ref{Fig1:neutron_maps_lowT} of the main text was obtained. Figure \ref{FigS2:additional_maps_highT} presents the measured INS intensity [panels (a)--(e)] and the corresponding line cuts [panels (f)--(j)] for the $1.7 < T < 1.9$~K data, from which the spectral function of Fig.~\ref{Fig2:neutron_maps_highT} of the main text was obtained. Here, the intensity is shown on a coarser scale to better visualize the smearing of the roton quasiparticle peak above 2$\Delta$. Figure \ref{FigS3:extended_maps_highT} presents extended data of Fig.~\ref{FigS2:additional_maps_highT} and Fig.~\ref{Fig2:neutron_maps_highT} of the main text, showing the spectral function [panels (a)--(h)] and the corresponding constant-$Q$ cuts of the measured INS intensity with fits [panels (i)--(q)] at 8 pressures studied in our experiment, as shown in Fig.~\ref{Fig0:PhaseDiagram} of the main text.

For a coherent quasiparticle with an infinite lifetime and dispersion relation $\epsilon (Q)$, the spectral function, $f_{\Gamma = 0}(Q, E)$, is Dirac's delta-function of zero width, $f_{\Gamma = 0}(Q, E) = \delta(E - \epsilon(Q))$. Intensity measured in experiment is determined by the convolution of the spectral function with the instrument resolution function, $R(E)$, which is approximately Gaussian function of energy. For delta-function spectral function, the resulting intensity distribution is a Gaussian peak describing the instrument resolution and tracking the dispersion, $\epsilon(q)$, as in Fig.~\ref{Fig1:neutron_maps_lowT} of the main text and Fig.~\ref{FigS1:additional_maps_lowT}. Or, for elastic scattering from the sample can, positioned at $E = 0$.

All data processing, including histogramming, fitting, and figure preparation, were done using custom Python scripts.

\subsection{Data analysis and peak fitting}
\label{Analysis}
To account for the finite lifetime of a quasiparticle, $\tau(Q) = \hbar/\Gamma(Q)$, which can also be wave vector dependent, we use the spectral function of a damped harmonic oscillator (DHO). The corresponding DHO dynamical susceptibility, $\chi''_{DHO} (E) = \frac{2 \Gamma(Q) \epsilon}{\left(\epsilon^2 - (\epsilon^2(Q)+\Gamma^2(Q)) \right)^2 + (2 \Gamma(Q) \epsilon)^2}$, is related to the dynamical structure factor measured by INS via fluctuation-dissipation theorem, which simply divides $\chi''_{DHO} (E)$ with a temperature-dependent detailed balance factor, $\left(1 - e^{-E/k_BT} \right)$, where $T$ is temperature and $k_B$ is Boltzmann constant \cite{ZaliznyakTranquada_2014,Zaliznyak_PRB1994}. In an underdamped case, $\Gamma(Q) < \epsilon(Q)$, $\chi''_{DHO} (E)$ can be represented as a difference of two Lorentzian functions positioned at $\pm \epsilon(Q)$ \cite{Zaliznyak_PRB1994}.
Convolution with a Gaussian resolution function replaces Lorentzians with Voigt functions, $V(\sigma, \Gamma(Q), E-\epsilon(Q))$, where Gaussian $\sigma$ is determined by the instrument energy resolution. By fitting elastic incoherent scattering from an empty sample can, we obtained 
energy resolution FWHM $\approx 0.09$~meV. The corresponding resolution-corrected DHO spectral function is,
\begin{equation}
\tilde{f}_{DHO}(Q, E) = \frac{A(Q)}{\left(1 - e^{-E/k_BT} \right)} \left( V(\sigma, \Gamma(Q), E-\epsilon(Q)) - V(\sigma, \Gamma(Q), E+\epsilon(Q)) \right) ,
\label{f_DHO_Voigt}
\end{equation}
where the factor $A(Q)$ ensures that the spectral function is normalized to 1, $\int \tilde{f}_{DHO}(Q, E) dE = 1$.

In our data analysis, we fit the constant-$Q$ cuts of the measured intensity, $I(Q, E)$, to the normalized spectral function of Eq.~\ref{f_DHO_Voigt} weighted by an intensity prefactor, $I(Q) \sim S(Q)$, which is refined together with $\epsilon(Q)$ and $\Gamma(Q)$. These fits are shown in the right panels, (f)--(j), of Figs.~\ref{FigS1:additional_maps_lowT} and \ref{FigS2:additional_maps_highT} and (i)--(q) of Fig.~\ref{FigS3:extended_maps_highT}. The corresponding fitted spectral functions, along with the measured ones, are shown in Figs.~\ref{Fig1:neutron_maps_lowT} and \ref{Fig2:neutron_maps_highT} of the main text. The resulting fitted peak parameters, the quasiparticle dispersion, $\epsilon(Q)$, intensity, $I(Q)$, and the width, $\Gamma(Q)$, are presented in Figures~\ref{FigS5:fits_lowT}--\ref{FigS7:extended_fits_highT}; $\Gamma(Q)$ is also presented in Figs.~\ref{Fig3:Gamma} and \ref{FigS8:Gamma_extended}.
For the low-temperature data (Figs.~\ref{Fig1:neutron_maps_lowT}, \ref{FigS1:additional_maps_lowT}, and \ref{FigS5:fits_lowT}) we also performed two-component fits accounting for a small intensity of multiparticle states above $E = 2\Delta$. For that, we added a second component in the form of a broad DHO with the lineshape determined from fitting intensity at the same $Q$ obtained in a supplementary measurement with $E_i = 11.22$~meV and only adjusting its amplitude in the fit (see Fig.~\ref{FigS4:DirectAnalysis}). The corresponding two-component fits are in good agreement with the single component fits and with the direct analysis (see below).
For the high-temperature data of Figs.~\ref{Fig2:neutron_maps_highT}, \ref{FigS2:additional_maps_highT}, and \ref{FigS3:extended_maps_highT} outside of the breakdown region, where the quasiparticle and the two-roton continuum do not overlap, we have also performed two-component fitting including the DHO spectral function of Eq.~\ref{f_DHO_Voigt} plus an Erf function modelling multiparticle states at higher energy above the $2 \Delta$ threshold. The resulting peak parameters are in good agreement with the single-component fits using Eq.~\ref{f_DHO_Voigt}; the corresponding $\epsilon(Q)$ is shown by filled symbols in Fig.~\ref{FigS7:extended_fits_highT}.

We have also performed direct analysis (DA) of the measured constant-$Q$ intensity profiles, which allows to evaluate effective Gaussian parameters of a peak from the raw data, without fitting (Fig.~\ref{FigS4:DirectAnalysis}). For that, we have evaluated a sloping background (BG) by connecting the average intensity points within a window of $1\times$FWHM of the peak and centered at a distance of $1.5\times$FWHM from the intensity maximum position, $E_0$, on each side of the peak (solid line segments and magenta boxes in Fig.~\ref{FigS4:DirectAnalysis}). Upon subtracting the resulting linear sloping BG intensity, $I_{BG}$, the integral intensity, $I_{int}(Q)$, the intensity-weighted (center-of-mass) peak position, $E_c(Q)$, and the mean square deviation of the intensity distribution within the peak, which gives the Gaussian $\sigma = $FWHM$/(\sqrt{8\ln{2}})$, were evaluated by direct numerical summation,
\begin{align}
I_{int}(Q) = \sum_{j}{\left( I(Q,E_j) - I_{BG} \right) \Delta E_j} \, , \\
E_c(Q) =  \frac{1}{I_{int}} \sum_{j}{E_j \left( I(Q,E_j) - I_{BG} \right) \Delta E_j} \, , \\
\sigma(Q) =\sqrt{\frac{1}{I_{int}} \sum_{j}{ \left( E_j - E_c(Q) \right)^2 \left( I(Q,E_j) - I_{BG} \right) \Delta E_j} } \, .
\label{DA}
\end{align}
Gaussian profiles with so determined parameters are shown by solid lines in Fig.~\ref{FigS4:DirectAnalysis}. In Fig.~\ref{FigS5:fits_lowT}, (a)--(e), the resulting peak position, which tracks the quasiaprticle dispersion (red triangles), and FWHM (small triangles at the bottom), are shown. The peak position determined from DA is in perfect agreement with the fitted peak position shown on top. The peak integral intensity obtained from the above DA is shown by red triangles in Fig.~\ref{FigS5:fits_lowT}, (f)--(j).

\subsection{Quasiparticle dispersion fitting}
\label{Bogolyubovdispersion}
Following Feynman \cite{Feynman_PhysRev1954} and Bohm and Salt \cite{BohmSalt_RMP1967}, we assume that for describing collective excitations in liquid helium at low energy, the strong inter-atomic interactions which create backflow when helium atoms move can be accounted for by introducing an effective mass, $m^\star = m\cdot m_{^4He}$ ($m_{^4He}$ is the nominal atomic mass of helium atom, and $m$ is the relative effective mass; a na{\"i}ve estimate for a sphere moving in a liquid of the same density gives $m = 1.5$ \cite{Feynman_PhysRev1954}). We further assume that the residual interaction between these effective-mass Bose particles representing helium atoms in superfluid helium can be approximated by a ``soft sphere'' potential, $V(r) = V_0 \Theta (a_0 - r)$ \cite{BohmSalt_RMP1967}, which imposes a constant energy cost, $V_0$, for two helium atoms to be within a distance $a_0$ from each other and is zero otherwise ($\Theta (x)$ is the Heaviside function), plus a weaker, higher-order interactions. Hence, we describe the dispersion of phonon-roton quasiparticles using the Bogolyubov expression for a weakly interacting Bose gas with an effective Hamiltonian with mass $m^\star$ and an effective theta-function interaction potential, $V_0 \Theta (a_0-r)$ \cite{Feynman_PhysRev1954,BogolyubovZubarev_JETP1955,BohmSalt_RMP1967,Sunakawa_ProgTheorPhys1969,IsiharaSamulski_PRB1977,ZaliznyakTranquada_2014},
\begin{equation}
\label{E_q_Bogolyubov}
\epsilon_0(Q) = \frac{\hbar^2}{m^\star a_0^2}\sqrt{(Q a_0)^4 + 8\pi v_0 \frac{m^\star a_0^2}{\hbar^2} \left(\frac{\sin(Q a_0)}{Q a_0} - \cos(Q a_0)\right)} .
\end{equation}
Here, $v_0 =V_0 n a_0^3 $, $n$ is the number density of helium atoms, and $\hbar^2/m^\star \approx 0.523/m$~meV$\cdot$~\AA$^2$. The above Bogolyubov dispersion,  Eq.~\eqref{E_q_Bogolyubov}, adequately reproduces phonon-maxon-roton behavior with a pronounced roton minimum at $Q_r = Q_r(m, a, v_0)$ and $\epsilon_0 (Q_r) = \Delta(m, a_0, v_0)$ and with a nearly quadratic dispersion in its vicinity\cite{Sunakawa_ProgTheorPhys1969,IsiharaSamulski_PRB1977,ZaliznyakTranquada_2014}. The energy and the wavevector position of the roton minimum are determined by the parameters of the effective Hamiltonian, $m$, $a_0$, and $V_0$.

When the dispersion of the Bogolyubov quasiparticles, Eq.~\eqref{E_q_Bogolyubov}, approaches the threshold of two-roton continuum, $2\Delta$, the quasiparticle energy, $\epsilon_0(Q)$, can be modified significantly by the interactions of phonon-roton quasiparticles with their own continuum, as first pointed out by Pitaevskii. In the absence of such interactions, the single-particle dispersion, Eq.~\eqref{E_q_Bogolyubov}, is simply superimposed on multiparticle states, and in particular on the two-roton continuum of states existing above the threshold energy $2\Delta$. Accounting for the interaction, first yields a ``repulsion'' effect, which lowers the quasiparticle energy and leads to ``squaring'' of the measured dispersion in the maxon region, which is clearly seen in the experimental data of Figs.~\ref{Fig1:neutron_maps_lowT}, \ref{FigS1:additional_maps_lowT}, \ref{FigS5:fits_lowT} (see also Refs.~\cite{Montfrooij_arXiv2006,Godfrin_PRB2021}).

Since we do not account for these repulsion effects of quasiparticle interaction with the two-roton continuum in our dispersion fitting, we constrain the energy range of the experimental data used for fitting to $E \lesssim 0.8 \cdot 2\Delta$ (short-dashed horizontal line in Figs.~\ref{FigS5:fits_lowT}--\ref{FigS7:extended_fits_highT}), assuming that the interaction and the resulting repulsion effects can be neglected for these lower energies, sufficiently far from $2\Delta$. The dispersion ``squaring'' effect of repulsion becomes more pronounced at higher pressures, where the two-roton threshold energy decreases. Hence, with the increasing pressure we observe an increasing discrepancy between the fitted dispersion of Eq.~\eqref{E_q_Bogolyubov} and the experimental points in the maxon region (Figs.~\ref{Fig1:neutron_maps_lowT}, \ref{FigS1:additional_maps_lowT}, \ref{FigS5:fits_lowT}). Note that there is an observable repulsion effect even at ambient pressure (1 bar), which was not noted previously.

The dispersion parameters, $m$, $a_0$, and $v_0$, obtained from our dispersion fitting for different pressures and temperatures are summarized in Fig.~\ref{FigS8:Gamma_extended}(e). Consistent with what could be expected on physical grounds, both effective mass, $m$, and the interaction radius, $a_0$, are weakly pressure- and temperature-dependent (these are effective Hamiltonian parameters, which are determined by integrating out the high-energy physics). The effective mass (circles) stays within few percent of $m \approx 1.6$, very close to the na\"{i}ve estimate of 1.5 \cite{Feynman_PhysRev1954,BohmSalt_RMP1967}, and $a_0$ [squares and the right scale in Fig.~\ref{FigS8:Gamma_extended}(e)] remains within a couple percent of $a_0 \approx 2.65$~\AA, which is very close to where the realistic interaction potential of helium atoms crosses zero energy \cite{Sunakawa_ProgTheorPhys1969}. The effective interaction parameter, $v_0$ [triangles in Fig.~\ref{FigS8:Gamma_extended}(e)], on the other hand, moderately increases with the increasing pressure (and density), as could also be expected on physical grounds. The pressure and temperature dependence of the roton gap obtained from our fitting (circles) and from the raw data (triangles) is presented in Fig.~\ref{FigS8:Gamma_extended}(f).

\subsection{Landau-Khalatnikov expression for quasiparticle damping}
\label{LK}
The LK expression \cite{LandauLifshitz_Vol9_1980,Haar_Landau_collected_papers_book1965,Khalatnikov_book,Gibbs_JPCM1999,Fak_PRL2012} for the roton quasiparticle width, $\Gamma$, is
\begin{equation}
\Gamma (T) = \gamma_R \sqrt{T} \left( 1 + \alpha\sqrt{\mu T} \right) e^{- \Delta(T)/T} ,
\end{equation}
where we use the most recent experimentally determined values, $\mu = 0.144$, $\gamma_R = 49.6(11)$~K$^{1/2}$, and $\alpha\sqrt{\mu} =  0.0603$~K$^{1/2}$ \cite{Fak_PRL2012}. 

\begin{figure}[p!]
\includegraphics[width=0.8\textwidth]{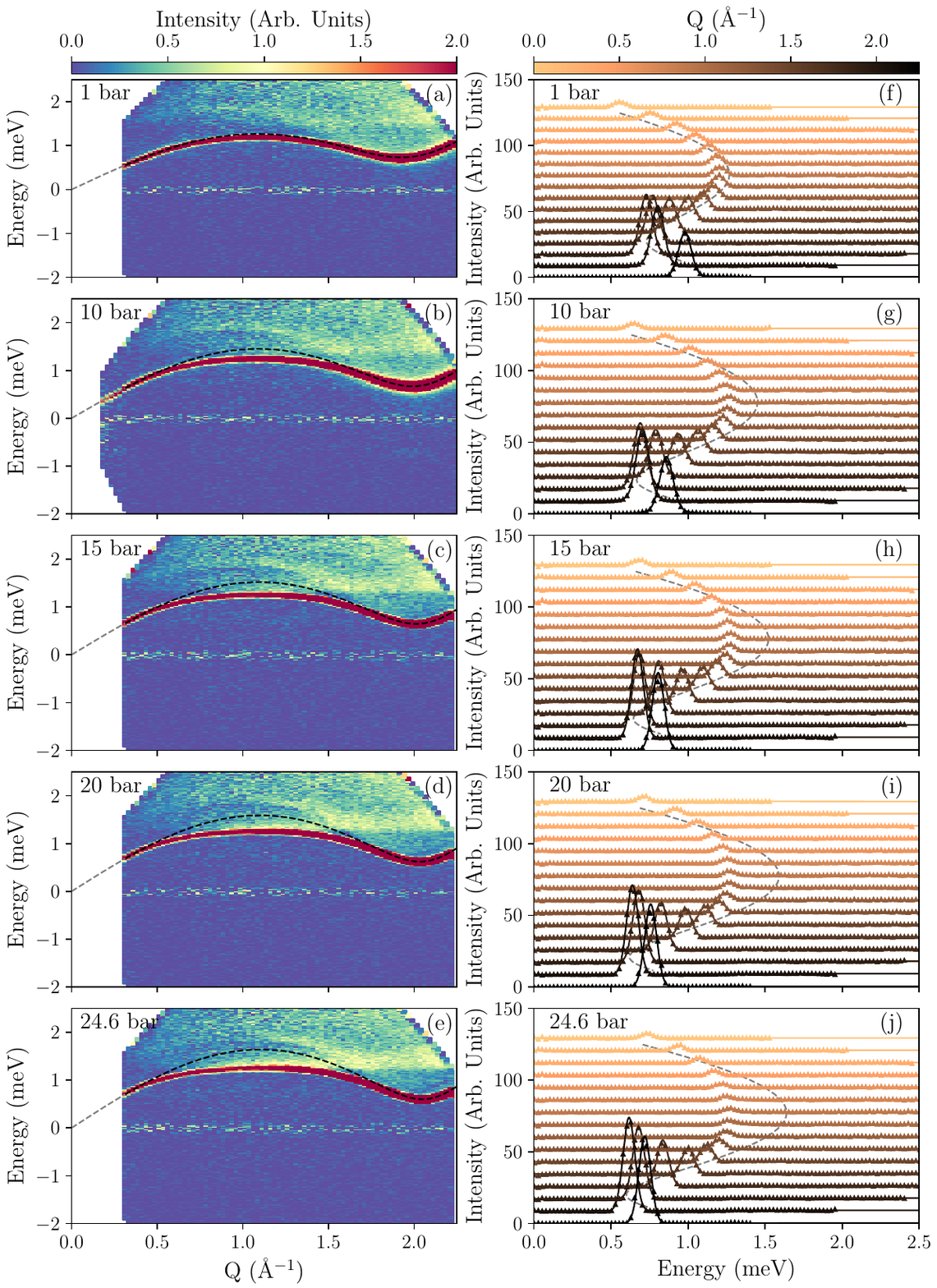}
\caption{{\bf INS intensity for $^4$He at $T < 0.9$~K}. (a)--(e) Color map of the measured neutron scattering intensity extending to negative energy transfers, which corresponds to the spectral function shown in Fig.~\ref{Fig1:neutron_maps_lowT} of the main text (note the accuracy of the empty can background subtraction near $E = 0$). (f)--(j) selected constant-$Q$ cuts of the corresponding measured INS intensity with fits to resolution-corrected DHO lineshape. The dashed curve is the fitted Bogolyubov dispersion \cite{Feynman_PhysRev1954,BogolyubovZubarev_JETP1955,BohmSalt_RMP1967,Sunakawa_ProgTheorPhys1969,IsiharaSamulski_PRB1977,ZaliznyakTranquada_2014} without account for multi-particle interactions, as in Figs.~\ref{Fig1:neutron_maps_lowT},~\ref{Fig2:neutron_maps_highT} of the main text. }
\label{FigS1:additional_maps_lowT}
\end{figure}

\begin{figure}[p!]
\includegraphics[width=0.8\textwidth]{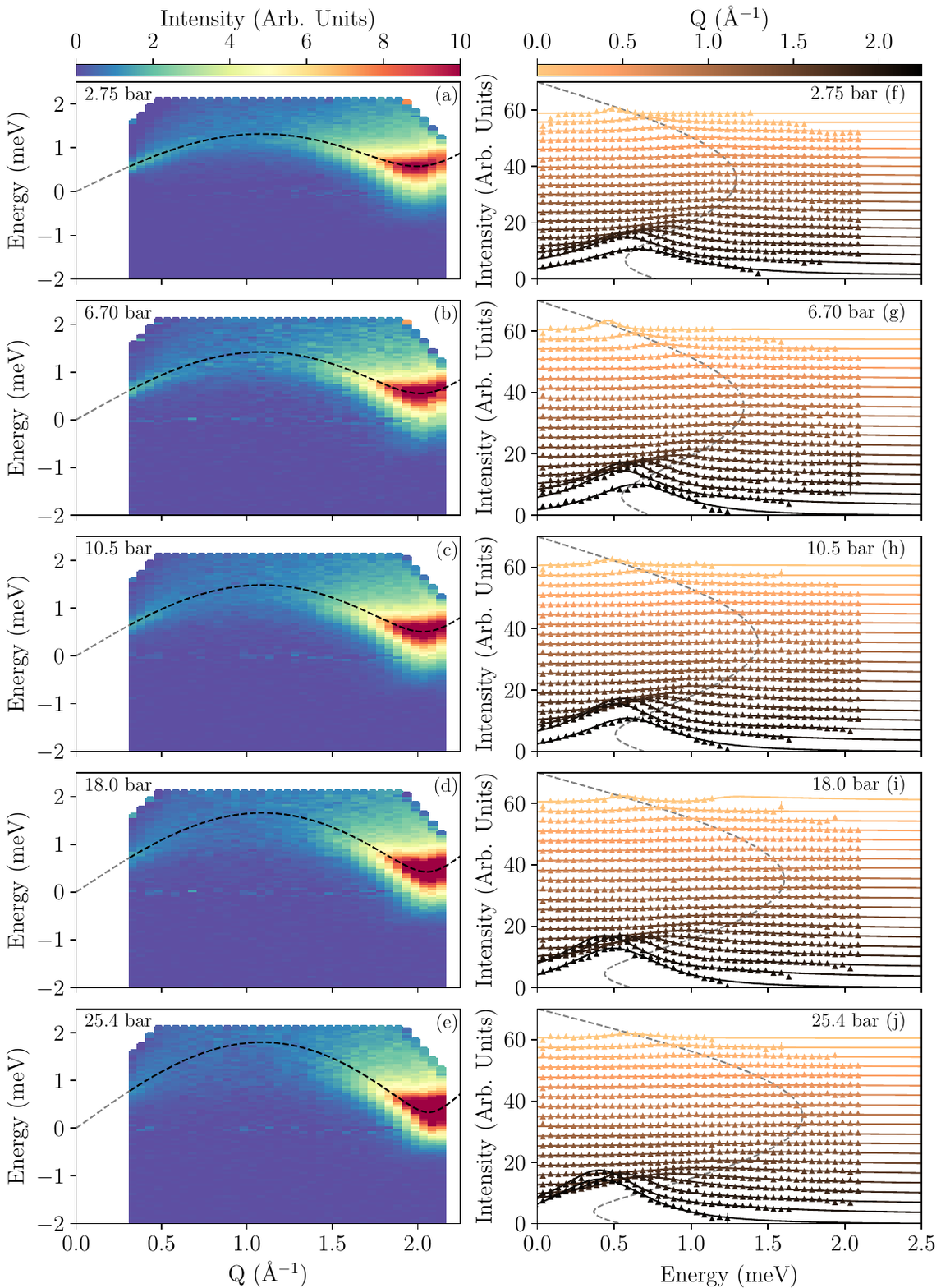}
\caption{{\bf INS intensity for $^4$He at $1.6 < T < 1.9$~K}. (a)--(e), color map of the measured neutron scattering intensity extending to  negative energy transfers, which corresponds to the spectral function shown in Fig.~\ref{Fig2:neutron_maps_highT} of the main text (note the accuracy of the empty can background subtraction near $E = 0$); intensity is shown on a coarser scale, [0,10], compared to Fig.~\ref{FigS1:additional_maps_lowT}, emphasizing the drop in roton intensity above 2$\Delta$. (f)--(j), selected constant-$Q$ cuts of the corresponding measured INS intensity with fits to resolution-corrected DHO lineshape. The dashed curve is the fitted Bogolyubov dispersion \cite{Feynman_PhysRev1954,BogolyubovZubarev_JETP1955,BohmSalt_RMP1967,Sunakawa_ProgTheorPhys1969,IsiharaSamulski_PRB1977,ZaliznyakTranquada_2014} without account for multi-particle interactions, as in Figs.~\ref{Fig1:neutron_maps_lowT},~\ref{Fig2:neutron_maps_highT} of the main text. }
\label{FigS2:additional_maps_highT}
\end{figure}

\begin{figure}[p!]
\includegraphics[width=0.8\textwidth]{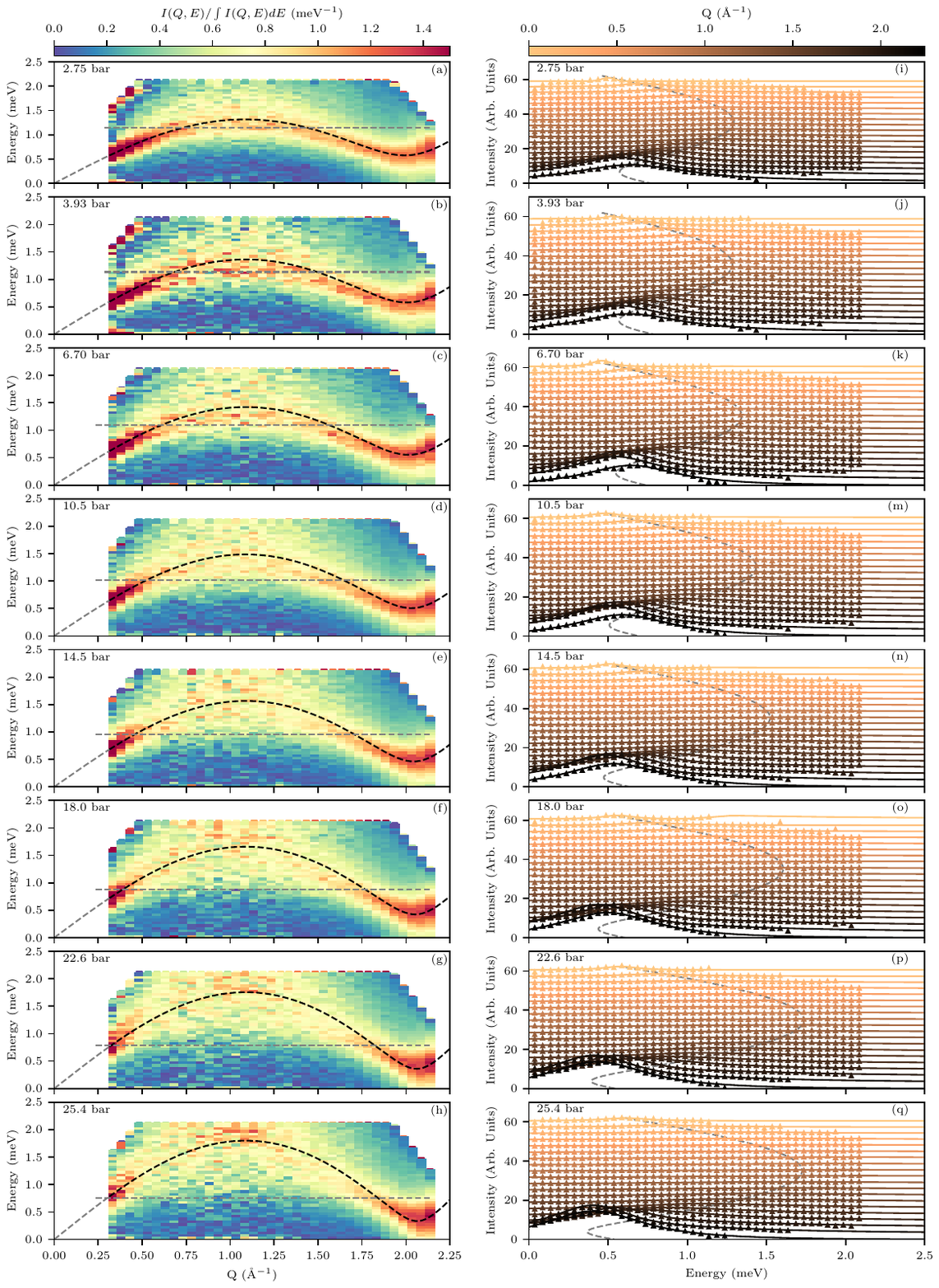}
\caption{{\bf Extended Figure~\ref{Fig2:neutron_maps_highT} including 8 measured pressures. Breakdown of phonon-roton sound-wave quasiparticle near the crystallization and superfluid transition.} (a)--(h) color contour maps of the spectral density of neutron scattering intensity at pressures $P = $~2.75(3), 3.93(3), 6.76(8), 10.50(3), 14.46(3), 18.03(3), 22.4(4), and 25.36(3) bar (top to bottom), at temperatures in $1.7$~K to $1.9$~K range, as shown in Fig.~\ref{Fig0:PhaseDiagram}. As in Figs.~\ref{Fig1:neutron_maps_lowT},~\ref{Fig2:neutron_maps_highT}, the dashed curve is the fitted Bogolyubov dispersion without account for multi-particle interactions, the horizontal dashed line marks the decay threshold energy, $2\Delta$. (i)--(q) selected constant-$Q$ cuts of the corresponding measured neutron scattering intensity with fits to resolution-corrected DHO lineshape. The substantial DHO width reflects short quasiparticle lifetime, which decreases dramatically above the decay threshold where the spectral density becomes extremely blurred.}
\label{FigS3:extended_maps_highT}
\end{figure}

\begin{figure}[p!]
\includegraphics[width=0.8\textwidth]{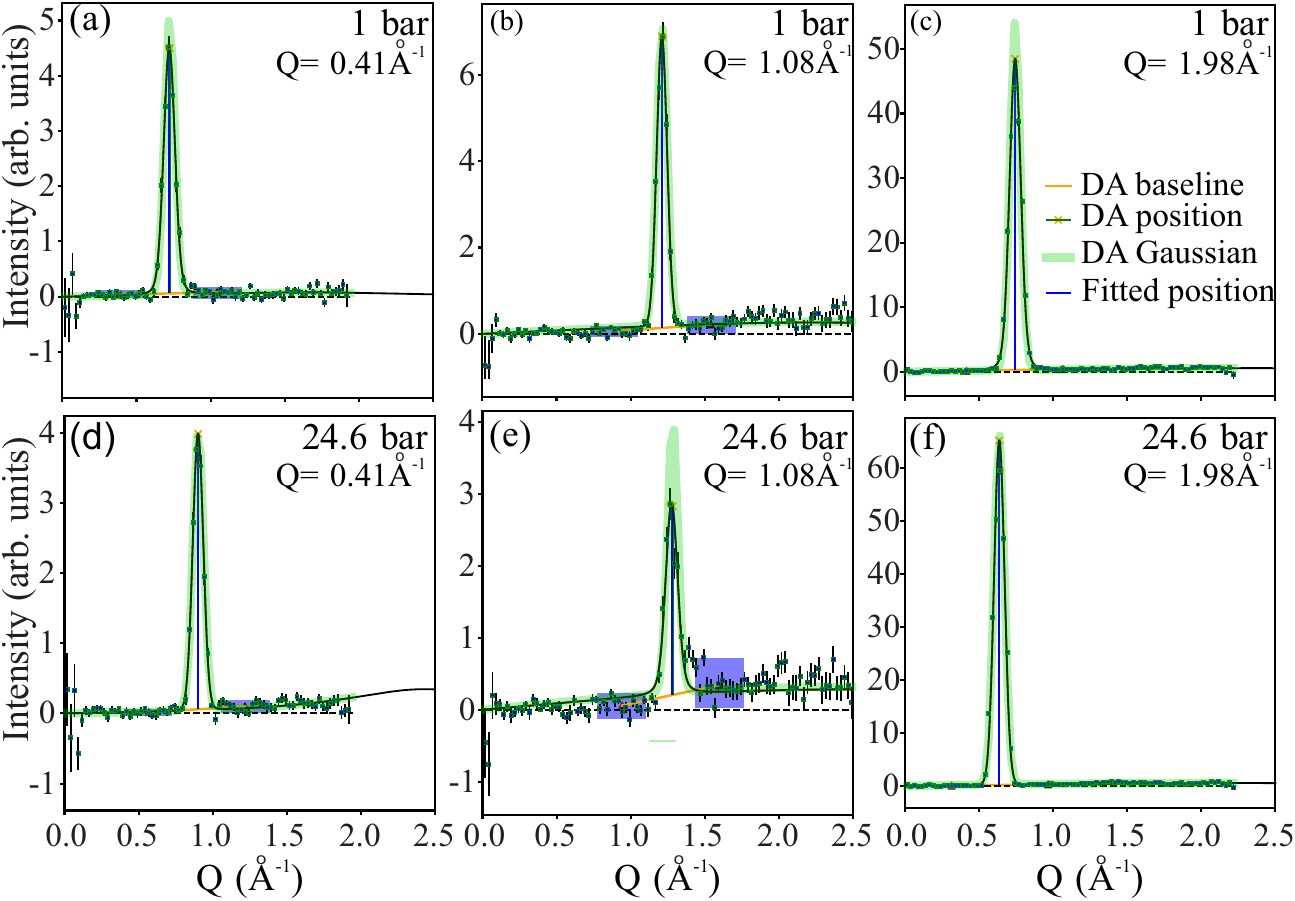}
\caption{{\bf Peak fit vs direct analysis.} (a)--(c) Points are the representative data showing measured neutron intensity for $P = 1.22$~bar and $T = 0.86$~K for wave vectors $0.41$~\AA$\color{blue} ^{-1}$, $1.08$~\AA$\color{blue} ^{-1}$, $1.08$, and $1.98$~\AA$\color{blue} ^{-1}$, from (a) to (c); (d)--(f) similar data at $P = 24.57$~bar and $T = 0.37$~K. Error bars show one standard deviation. In all panels, thin black line shows the two-component fit using Eq.~\ref{f_DHO_Voigt} as described in the text (vertical line is a fitted peak position). Thick light-green line is a Gaussian profile obtained using the peak position (shown by cross), intensity, and FWHM evaluated from direct analysis of the measured intensity as described in the text. Sloped segment under the peak shows the sloping background subtracted in direct analysis, which was evaluated by averaging points in purple rectangular regions, outside 2 FWHM from peak center. Error bars indicate one standard deviation.}
\label{FigS4:DirectAnalysis}
\end{figure}

\begin{figure}[p!]
\includegraphics[width=0.6\textwidth]{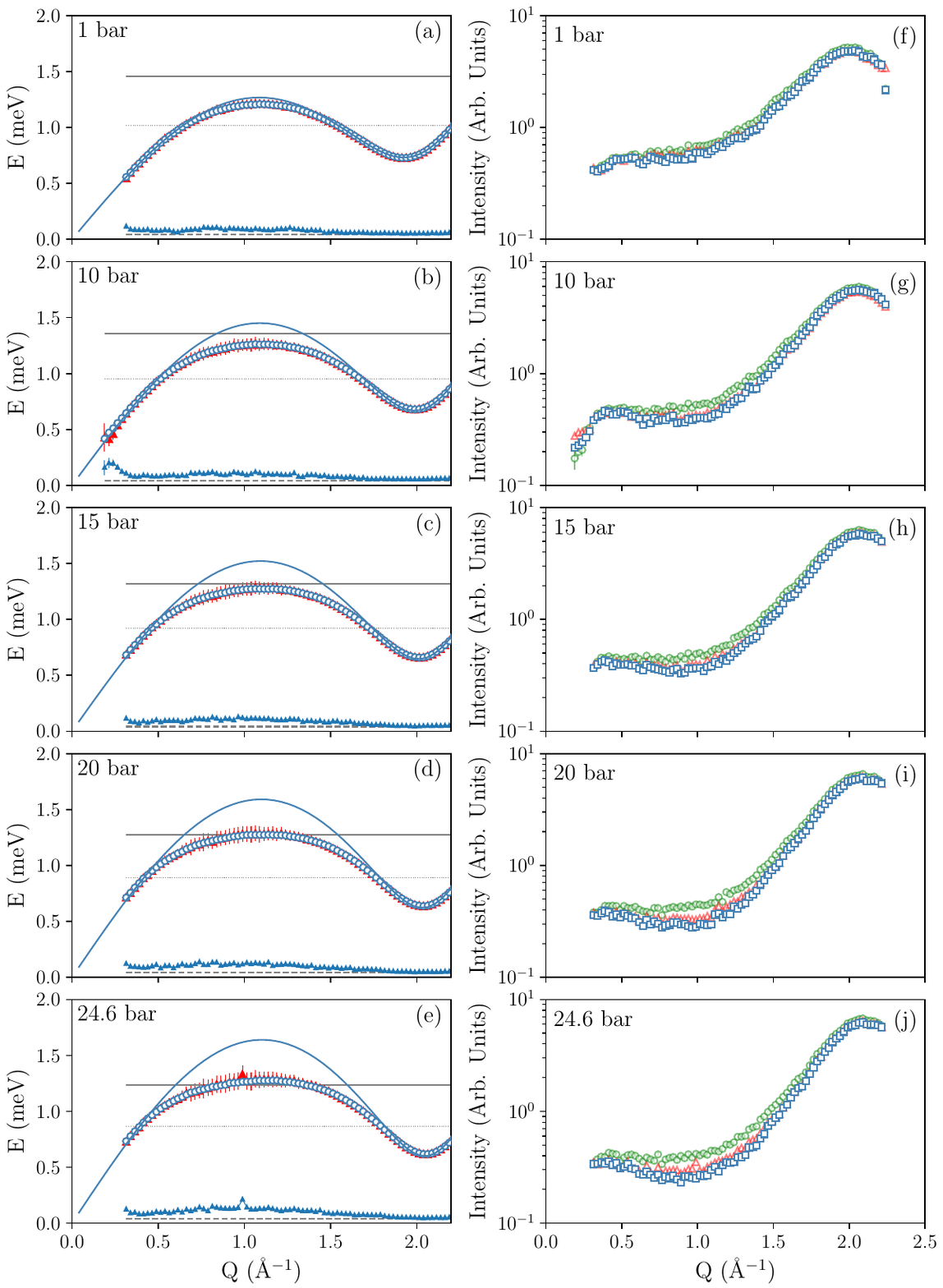}
\caption{{\bf The measured phonon-roton energy at $T < 0.9$~K with the fitted dispersion of Bogolyubov quasiparticles without multi-particle interactions}. (a)--(e) open circles show the bare undamped (Lorentzian center) energy, $\epsilon(Q)$, obtained from the DHO fits of the measured INS spectra in Fig.~\ref{Fig1:neutron_maps_lowT},~(f)--(j). The open red triangles (underneath the superimposed circles) with visible error bars show peak position obtained from direct analysis (DA) of the center-of-mass peak position in the measured intensity (where not visible, error bars are smaller than the symbol size). The solid horizontal line is the two-roton threshold energy, $2\Delta$; the short-dashed line shows the cutoff energy of the data used for the Bogolyubov dispersion fitting (only data below this line were used, assuming that interaction with the two-roton continuum in this range can be neglected). The long-dashed line at the bottom shows the energy resolution obtained from Gaussian fit of elastic incoherent scattering from the empty sample can; small triangles show width of the quasiparticle peak obtained from DA of the raw data, as a weighted mean-square deviation of the measured intensity from its center-of-mass peak position. (f)--(j) the integral intensity of the quasiparticle peak from fit (squares), DA (red triangles), and the measured intensity integrated in the $[-2\Delta, 2\Delta]$ range (green circles). Error bars indicate one standard deviation.}
\label{FigS5:fits_lowT}
\end{figure}

\begin{figure}[p!]
\includegraphics[width=0.6\textwidth]{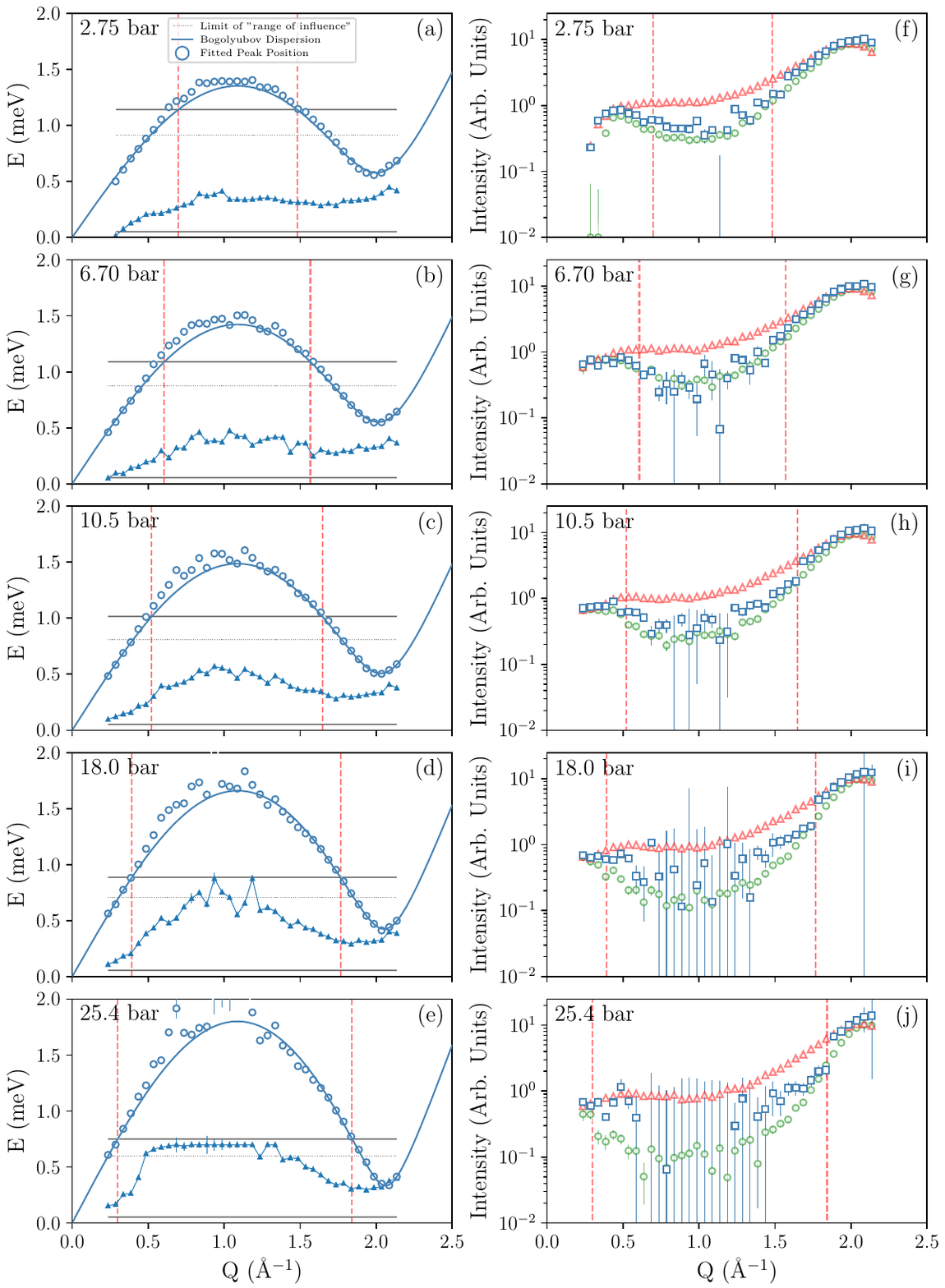}
\caption{{\bf The measured phonon-roton energy at $1.6 < T < 1.9$~K with the fitted dispersion of Bogolyubov quasiparticles without multi-particle interactions}. (a)--(e) open circles show the bare undamped (Lorentzian center) energy, $\epsilon(Q)$, obtained from single-component DHO fits of the measured INS spectra in Fig.~\ref{Fig2:neutron_maps_highT},~(f)--(j). Where not visible, error bars are smaller than the symbol size. The horizontal lines are same as in Fig.~\ref{FigS5:fits_lowT}, the two-roton threshold energy, $2\Delta$ (solid), the cutoff energy of the data used for the Bogolyubov dispersion fitting (short-dashed), and the energy resolution obtained from Gaussian fit of elastic incoherent scattering from the empty sample can (long-dashed). Small triangles show Lorentzian HWHM of the quasiparticle peak, $\Gamma$ (shown in Figs.~\ref{Fig3:Gamma},~\ref{FigS8:Gamma_extended}), obtained from the DHO fit. (f)--(j) the integral intensity of the quasiparticle peak from fit (squares), DA (green circles), and the measured intensity integrated in the $[-2\Delta, 2\Delta]$ range (red triangles). The vertical dashed lines show the region of momenta, $[Q_{c 1}, Q_{c 2}]$, where $\epsilon(Q) < 2\Delta$ and decays are allowed for non-interacting quasiparticles (solid symbols in Figs.~\ref{Fig3:Gamma},~\ref{FigS8:Gamma_extended}). While peak position in the breakdown range, $[Q_{c 1}, Q_{c 2}]$, might not correspond to a quasiparticle, it shows that maximum intensity within the two-roton continuum occurs near the non-interacting quasiparticle dispersion. Error bars indicate one standard deviation.}
\label{FigS6:fits_highT}
\end{figure}

\begin{figure}[p!]
\includegraphics[width=0.8\textwidth]{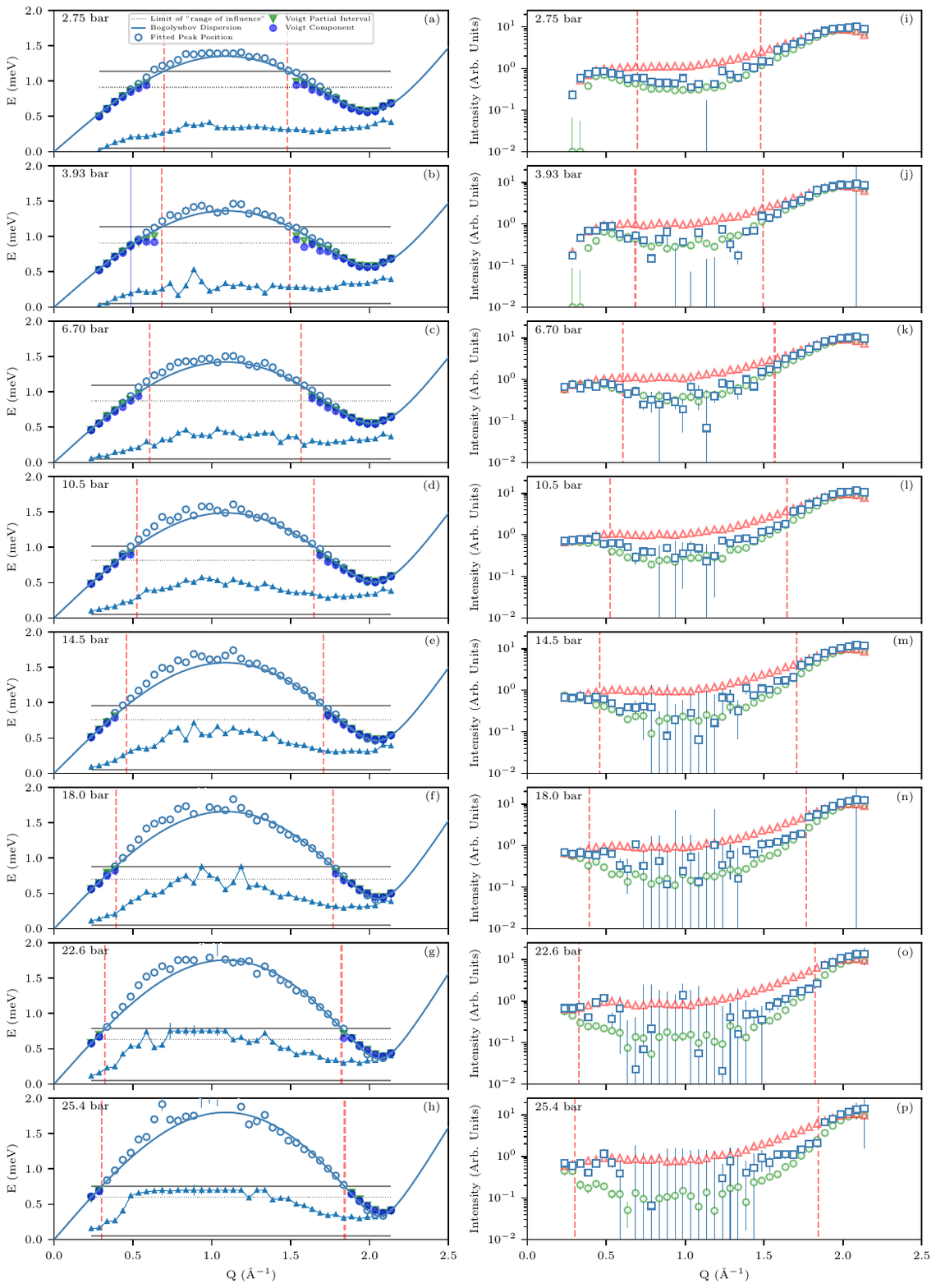}
\caption{{\bf Extended Figure~\ref{FigS6:fits_highT} including 8 measured pressures. The measured phonon-roton energy at $1.6 < T < 1.9$~K with the fitted dispersion of Bogolyubov quasiparticles without multi-particle interactions}. (a)--(h) in addition to symbols shown in Figure~\ref{FigS6:fits_highT}~(a)--(e), outside the breakdown range, $[Q_{c 1}, Q_{c 2}]$, the DHO position obtained by fitting data in a limited energy range below $2\Delta$ (green triangles) and obtained from two-component fits including a DHO and a broad signal modelled by an Erf function accounting for multi-particle states at higher energy above $2 \Delta$ (filled symbols) are also shown. For all three fits, the positions are in good agreement with each other and yield similar (within error bar) values of the Bogolyubov dispersion parameters (Fig.~\ref{FigS8:Gamma_extended}~(e),(f)). The lines and symbols in (i)--(p) are same as in Fig.~\ref{FigS6:fits_highT}~(f)--(j). Error bars indicate one standard deviation.}
\label{FigS7:extended_fits_highT}
\end{figure}

\begin{figure*}[p!]
\includegraphics[width=1.\textwidth]{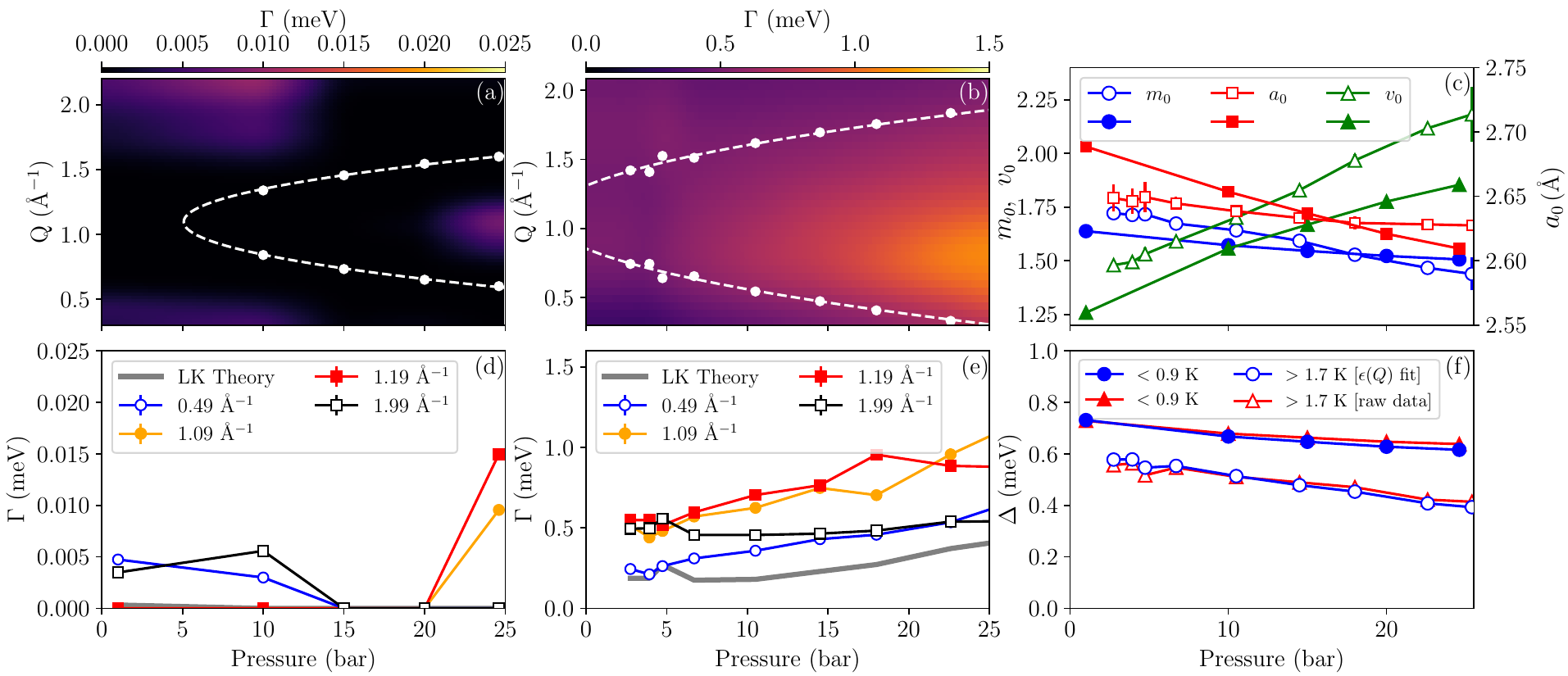}
\caption{{\bf Extended figure~\ref{Fig3:Gamma}. The quasiparticle width and dispersion parameters.} (a)--(d) are same as in Fig.~\ref{Fig3:Gamma}.
Color contour map of the DHO half width at half maximum (HWHM), $\Gamma$, which parameterizes the quasiparticle lifetime, $\tau \sim h/\Gamma$, obtained by interpolation of the fit results as a function of pressure, (a) for the low-temperature data of Fig.~\ref{Fig1:neutron_maps_lowT} and (c) for the data of Fig.~\ref{Fig2:neutron_maps_highT} with pronounced decays. In (a), for the low-pressure data measured at $0.87(1)$~K, small width reflecting finite lifetime effects due to collisions can be seen for some wave vectors, while for the $0.35(2)$~K data at higher $P \gtrsim 20$~bar the lifetime effects due to decays become apparent.
The solid symbols with the parabolic fit (dashed line) show the boundary of the pressure-dependent quasiparticle breakdown region of momenta, $[Q_{c 1}, Q_{c 2}]$, where decays are expected for non-interacting quasiparticles with the fitted Bogolyubov dispersion \cite{Feynman_PhysRev1954,BogolyubovZubarev_JETP1955,BohmSalt_RMP1967,Sunakawa_ProgTheorPhys1969,IsiharaSamulski_PRB1977,ZaliznyakTranquada_2014} of Figs.~\ref{Fig1:neutron_maps_lowT},~\ref{Fig2:neutron_maps_highT}(a)--(e) (see also Figs.~\ref{FigS1:additional_maps_lowT}--\ref{FigS7:extended_fits_highT}). (b) The pressure dependence of $\Gamma$ for typical wave vectors in the phonon (open circles), maxon (filled circles and squares) and roton (open squares) regions for the data in (a) and (d) for the data in (c). The grey line shows $\Gamma$ obtained from LK theory \cite{Andersen_PRL1996,Fak_PRL2012}. The experimental $\Gamma$ obtained using a single-component DHO fit is somewhat over-estimated by inclusion of multi-particle states in the fitted intensity, however, its variation with pressure and $Q$ adequately exposes the physics of quasiaprticle decays.
(e) The pressure dependence of the Bogolyubov dispersion parameters, the effective mass $m_0$ (in units of $m_{^4He}$, circles), the soft-core potential radius, $a_0$ (squares), and the effective potential strength, $v_0$ (triangles), for the low-temperature data of Figs.~\ref{Fig1:neutron_maps_lowT},~\ref{FigS1:additional_maps_lowT},~\ref{FigS5:fits_lowT} (filled symbols) and for the $1.6 < T < 1.9$~K data of Figs.~\ref{Fig2:neutron_maps_highT},~\ref{FigS3:extended_maps_highT},~\ref{FigS7:extended_fits_highT} (open symbols). (f) The pressure dependence of the roton gap, $\Delta$, determined from the fitted dispersion (circles) and from the raw data (triangles) for the low-T data (filled symbols) and for the $1.6 < T < 1.9$~K data (open symbols). Error bars indicate one standard deviation and where not visible are smaller than the symbol size.} 
\label{FigS8:Gamma_extended}
\end{figure*}

\end{widetext}

\end{document}